\documentclass[prb,twocolumn,preprintnumbers,amsmath,amssymb]{revtex4}

\usepackage{graphicx}
\usepackage{amssymb,amsmath}
\usepackage{dcolumn}
\usepackage{bm}
\usepackage{color}
\usepackage{enumerate}
\usepackage{epstopdf}

\begin{document}

\title{Electronic states bound by repulsive potentials in graphene irradiated by a circularly polarized electromagnetic field}

\author{O. V. Kibis$^1$}\email{oleg.kibis(c)nstu.ru}
\author{M. V. Boev$^1$}
\author{I. V. Iorsh$^{1,2}$}
\author{V. M. Kovalev$^1$}

\affiliation{$^1$Department of Applied and Theoretical Physics,
Novosibirsk State Technical University, Karl Marx Avenue 20,
Novosibirsk 630073, Russia}
\affiliation{$^2$Department of Physics, Queen's University, Kingston, Canada}

\begin{abstract}
In the framework of the Floquet theory of periodically driven quantum systems, it is demonstrated that irradiation of graphene by a circularly polarized electromagnetic field induces an attractive area in the core of repulsive potentials. Consequently, the quasi-stationary electron states bound by the repulsive potentials appear. The difference between such field-induced states in graphene and usual systems with the parabolic dispersion of electrons is discussed and possible manifestations of these states in electronic transport and optical spectra of graphene are considered.
\end{abstract}
\pacs{}

\maketitle

\section{Introduction}
The control of conduction electrons in condensed-matter structures by a high-frequency off-resonant electromagnetic field --- also known as the Floquet engineering --- has become an excited area of modern physics~\cite{Oka_2019,Basov_2017,Goldman_2014,Bukov_2015,Eckardt_2015,
Casas_2001,Nuske_2020,Liu_2022,Seshadri_2022,
Kobayashi_2023}. If a field frequency much exceeds characteristic electron frequencies, the field cannot be absorbed by electrons but only dresses them. As a result, such a dressing field substantially modifies electronic properties of various nanostructures, including
quantum rings~\cite{Koshelev_2015,Kozin_2018}, quantum wells~\cite{Lindner_2011}, topological insulators~\cite{Rechtsman_2013,Wang_2013,Zhu_2023,Torres_2014}, Moir\'e structures~\cite{Topp_2019,Vega_2020,Anderson_2020}, materials with tilted Dirac cones~\cite{Champo_2019,Sierra_2019,Iurov_2022}, graphene~\cite{Oka_2009,Kibis_2010,Kristinsson_2016,Syzranov_2013,Usaj_2014,Perez_2014,Cavalleri_2020,Iurov_2022_1} and related two-dimensional materials~\cite{Sie_2015,Iurov_2019,Iurov_2020}, etc.
Particularly, it was found recently that a circularly polarized electromagnetic field can induce an attractive area in the core of two-dimensional (2D) repulsive potentials~\cite{Kibis_2019}. This results in the quasi-stationary electron states bound by the repulsive potentials, which substantially modify electronic properties of irradiated 2D systems~\cite{Kibis_2020,Kibis_2021_1,Iorsh_2021}. However, the theory of these states was developed before only for 2D systems with the parabolic dispersion of electrons. The present article is aimed to extend the theory over 2D systems with the linear (Dirac) electron dispersion which takes place in graphene and related materials.

The article is organized as follows. In Sec.~II, the model of the discussed effect is developed. In Sec.~III, possible manifestations of the optically induced quasi-stationary electron states bound by repulsive potentials in transport and optical phenomena are discussed. The last two sections contain the conclusion and acknowledgements.

\section{Model}
\begin{figure}[!ht]
\includegraphics[width=1.\columnwidth]{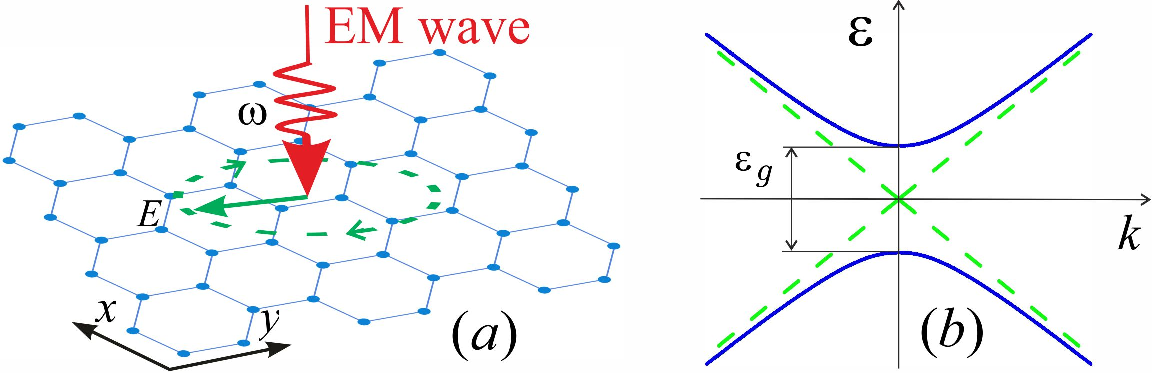}
\caption{(a) Sketch of the system under consideration: Graphene irradiated by a circularly polarized electromagnetic wave with the frequency $\omega$ and the electric field amplitude $E$; (b) Energy band structure of graphene near the Dirac point in the presence of circularly polarized irradiation (solid
lines) and in the absence of one (dashed lines).}\label{Fig.1}
\end{figure}
Let us consider a graphene sheet which lies in the plane $(x,y)$
and is subjected to a circularly polarized electromagnetic wave propagating perpendicularly to the plane (see Fig.~1a). Then electronic properties of the irradiated graphene sheet near its Dirac point are described by the Hamiltonian~\cite{Kibis_2010,Kristinsson_2016}
\begin{equation}\label{AH0}
\hat{\cal{H}}=v_0\bm{\sigma}\cdot(\hbar\mathbf{k}-e\mathbf{A}(t)/c),
\end{equation}
where $\bm{\sigma}=(\sigma_x,\sigma_y)$ is the Pauli matrix
vector, $\mathbf{k}=(k_x,k_y)$ is the electron wave vector in the
graphene plane, $v_0$ is the electron velocity in graphene near the
Dirac point ($\mathbf{k}=0$), $e$ is the electron (hole) charge,
\begin{equation}\label{A}
\mathbf{A}(t)=(A_x,A_y)=\frac{cE}{\omega}\left(\cos\omega
t,\sin\omega t\right)
\end{equation}
is the vector potential of the electromagnetic wave, $E$ is the electric field amplitude of the wave, and
$\omega$ is the wave frequency. Solving the
Schr\"odinger problem with the periodically time-dependent Hamiltonian (\ref{AH0}), it was demonstrated earlier that the circularly polarized field \eqref{A} opens the band gap in the Dirac point~\cite{Kibis_2010,Kristinsson_2016},
\begin{equation}\label{Eg}
\varepsilon_g=\sqrt{\left({\hbar\omega}\right)^2+4W^2}-\hbar\omega,
\end{equation}
where $W={v_0|e|E}/{\omega}$ is the kinetic energy of electron (hole) rotation under the field (see Fig.~1b). In the following, we will assume that the field \eqref{A} is strong enough to satisfy the condition
\begin{equation}\label{c}
W\gg\hbar\omega.
\end{equation}
Under this condition, the interband  absorption of the field \eqref{A} is forbidden by the energy conservation law since the band gap \eqref{Eg} much exceeds the photon energy. To proceed, let us rewrite the Hamiltonian \eqref{AH0} in the matrix form as
\begin{align}\label{HD}
&\hat{\cal{H}}=\hbar v_0
\begin{pmatrix}
0 & k_x-ik_y\\
k_x+ik_y & 0
\end{pmatrix}\nonumber\\
&-\frac{v_0e}{c}
\begin{pmatrix}
0 & A_x-iA_y\\
A_x+iA_y & 0
\end{pmatrix}
&=
\begin{pmatrix}
0 & \varepsilon_\mathbf{k}e^{-i\theta}\\
\varepsilon_\mathbf{k}e^{i\theta}& 0
\end{pmatrix}
\end{align}
and apply the unitary transformation
\begin{equation}\label{U1}
\hat{\cal{U}}=\frac{1}{\sqrt{2}}
\begin{pmatrix}
e^{-i\theta/2} & e^{-i\theta/2}\\
e^{i\theta/2} & -e^{i\theta/2}
\end{pmatrix},
\end{equation}
where
\begin{equation}\label{Th}
\theta=\arctan\left(\frac{\hbar k_y-(eE/\omega)\sin\omega t}
{\hbar k_x-(eE/\omega)\cos\omega t}\right),
\end{equation}
and
\begin{equation}\label{ek}
\varepsilon_\mathbf{k}=v_0\sqrt{(\hbar k_x-eA_x/c)^2+(\hbar k_y-eA_y/c)^2}.
\end{equation}
Then the unitary transformed Hamiltonian \eqref{HD} reads
\begin{equation}\label{H1}
\hat{\cal H}^{\prime}= \hat{U}^\dagger\hat{\cal H}\hat{U} -i\hbar\hat{U}^\dagger\partial_t
\hat{U}=\begin{pmatrix}
\varepsilon_\mathbf{k} & f_\mathbf{k}\\
f_\mathbf{k} & -\varepsilon_\mathbf{k}
\end{pmatrix},
\end{equation}
where
\begin{equation}\label{f}
f_\mathbf{k}=\frac{\hbar\omega}{2}\left[\frac{(\hbar v_0^2e/c)\mathbf{k}\mathbf{A}(t)-W^2}{\varepsilon_\mathbf{k}^2}\right].
\end{equation}
It follows from Eq.~\eqref{ek} that the diagonal terms of the Hamiltonian \eqref{H1} are identical to the electron energy spectrum of graphene, $\varepsilon(\mathbf{k})=\pm\hbar v_0\sqrt{k_x^2+k_y^2}$, with the replacement $\hbar k_{x,y}\rightarrow\hbar k_{x,y}-eA_{x,y}/c$. Physically, they describe the time evolution of electron (hole) states within the conduction (valence) band of graphene under the field \eqref{A}. As to the non-diagonal terms \eqref{f}, they describe the mixing of the conduction and valence bands by the field. Under the condition \eqref{c}, these non-diagonal terms contribute to the electron states near the band edges with the smallness $\sim\hbar\omega/W$ and, therefore, can be omitted as a first approximation. Then the diagonal terms of the Hamiltonian \eqref{H1} yield the electron dispersion near the band edge ($\hbar v_0k\ll W$),
\begin{align}\label{ek0}
&\varepsilon_\pm(\mathbf{k})\approx \pm W \mp \frac{e}{|e|} \hbar v_{0} (k_{x}\cos\omega t + k_{y}\sin\omega t)\nonumber\\
&\pm \frac{\hbar^2v_0^2}{2 W} \left(k_{x}^2 \sin^2\omega t + k_{y}^2 \cos^2\omega t - k_{x} k_{y} \sin2\omega t\right),
\end{align}
where the signs ``$+$'' and ``$-$'' correspond to the conduction and valence band, respectively. Replacing the momentum $\hbar\mathbf{k}=(\hbar k_x,\hbar k_y)$ in Eq.~\eqref{ek0} with the momentum operator $\hat{\mathbf{p}}=(\hat{p}_x,\hat{p}_y)$, we arrive at the Hamiltonian
\begin{align}\label{H2}
&\hat{\cal H}_\pm=\pm W \mp \frac{e}{|e|} v_{0} (\hat{p}_x\cos\omega t + \hat{p}_y\sin\omega t)\nonumber\\
&\pm \frac{v_0^2}{2 W} \left(\hat{p}_x^2 \sin^2\omega t + \hat{p}_y^2 \cos^2\omega t - \hat{p}_x \hat{p}_y \sin2\omega t\right),
\end{align}
which describes the dynamics of charge carriers in the scatterer potential $U(\mathbf{r})$ near the band edge.

To remove the linear terms $\propto \hat{p}_{x,y}$ from the Hamiltonian \eqref{H2}, let us apply the Kramers-Henneberger (KH) unitary transformation~\cite{Kramers_1952,Henneberger_1968}. In the considered case of irradiated graphene, the KH transformation operator is
\begin{equation}\label{KH}
\hat{\cal U}_\pm=\exp\left\{\pm\frac{i}{\hbar}\int^{\,t}\left[
v_0\frac{e\mathbf{A}(\tau)}{|e\mathbf{A}(\tau)|}\hat{\mathbf{p}}-W\right]d\tau\right\}
\end{equation}
and, correspondingly, the unitary transformed Hamiltonian \eqref{H2} reads
\begin{align}\label{UH}
&\hat{\cal H}'_\pm=\hat{\cal U}^\dagger_\pm\hat{\cal H}_\pm\hat{\cal U}_\pm -
i\hbar\hat{\cal U}^\dagger_\pm\partial_t
\hat{\cal U}_\pm=U\big(\mathbf{r}-
\mathbf{r}_0(t)\big)\nonumber\\
&\pm \frac{v_0^2}{2 W} \left(\hat{p}_x^2 \sin^2\omega t + \hat{p}_y^2 \cos^2\omega t - \hat{p}_x \hat{p}_y \sin2\omega t\right),
\end{align}
where
\begin{equation}\label{rt}
\mathbf{r}_0(t)=r^\ast_0\left(-\sin\omega t,\cos\omega t\right)
\end{equation}
is the radius vector of circular trajectory of a charge carrier in graphene under the field \eqref{A}, and
\begin{equation}\label{r0}
{r}^\ast_0=\frac{v_0}{\omega}
\end{equation}
is the radius of the trajectory. It should be noted that the Hamiltonian \eqref{UH} with the periodically time-dependent potential is still physically identical to the Hamiltonian \eqref{H2}. Next, we need to make the high-frequency approximation. It follows from the conventional Floquet theory of periodically driven quantum systems that one can introduce the unitary transformation, $\hat{\cal U}_0(t)$, turning the time-dependent Hamiltonian \eqref{UH} into the effective stationary Hamiltonian,
\begin{equation}\label{HE}
\hat{\cal H}_{\mathrm{eff}}=\hat{\cal U}_0(t)^\dagger\hat{\cal H}'_\pm\hat{\cal U}_0(t) -
i\hbar\hat{\cal U}_0^\dagger(t)\partial_t
\hat{\cal U}_0(t),
\end{equation}
which can be written as an expansion in powers of $1/\omega$ (the Floquet-Magnus expansion)~\cite{Eckardt_2015,Goldman_2014,Bukov_2015,Casas_2001}, \begin{equation}\label{Hef}
\hat{\cal H}_{\mathrm{eff}}=\hat{\cal H}_{0}+\sum_{n=1}^\infty\frac{[\hat{\cal H}_{n},\hat{\cal H}_{-n}]}{n\hbar\omega}+{\it O}\left(\frac{1}{\omega^2}\right),
\end{equation}
where $\hat{\cal H}_{n}$ are the Fourier harmonics of the time-dependent Hamiltonian \eqref{UH}, $\hat{\cal H}'_\pm=\sum_{n=-\infty}^{\infty}\hat{\cal H}_{n}e^{in\omega t}$. In the high-frequency limit, one can restrict the expansion \eqref{Hef} by its main term,
\begin{equation}\label{H0}
\hat{\cal H}_0=\pm\frac{\hat{\mathbf{p}}^2}{2m^\ast_e}+U_0(\mathbf{r}),
\end{equation}
where
\begin{equation}\label{m}
m^\ast_e=\frac{2|e|E}{v_0\omega}
\end{equation}
is the field-induced effective mass of charge carriers in the irradiated graphene, and the time-averaged potential
\begin{equation}\label{U0}
U_0(\mathbf{r})=\frac{1}{2\pi}\int_{-\pi}^{\pi}U\big(\mathbf{r}-\mathbf{r}_0(t)\big)\,d(\omega
t)
\end{equation}
can be treated as a repulsive potential $U(\mathbf{r})$ dressed by the circularly polarized field \eqref{A}. In the following, the effective stationary Hamiltonian \eqref{H0} will be applied to describe electron dynamics in this potential.

\section{Results and discussion}
The Hamiltonian \eqref{H0} is equal to the Hamiltonian describing an irradiated 2D system with the parabolic electron dispersion if to make the replacements $m^\ast_e\rightarrow m_e$ and $r^\ast_0\rightarrow r_0$, where $m_e$ is the effective electron mass corresponding to the parabolic electron energy spectrum $\varepsilon(\mathbf{k})=\hbar^2\mathbf{k}^2/2m_e$ and $r_0=|e|E/m_e\omega^2$ is the radius of circular trajectory of an electron in the 2D system under the field \eqref{A} (see Ref.~\onlinecite{Kibis_2020}). Therefore, the behavior of electrons describing by these Hamiltonians is qualitatively similar. Particularly, it was demonstrated earlier that the dressed potential \eqref{U0} originated from an repulsive potential of most general form acquires the local minimum in its core, which leads to electron states confined near this minimum~\cite{Kibis_2019}. Restricting the analysis by the case of short-range scatterers and modelling their repulsive potential by the delta function, $U(\mathbf{r})=u_0\delta(\mathbf{r})$, the dressed potential \eqref{U0} reads~\cite{Kibis_2020}
\begin{equation}\label{U01}
U_0(\mathbf{r})=\frac{u_0\,\delta({r}-{r}^\ast_0)}{2\pi r^\ast_0},
\end{equation}
where $u_0>0$ signifies the strength of the repulsive potential, and $\mathbf{r}=(x,y)$ is the plane radius vector. Consequently, the dressing field given by Eq.~\eqref{A} transforms the repulsive delta potential into a ring-shaped delta potential barrier described by Eq.~\eqref{U01}. This transformation leads to the confinement of electron states within the enclosed region $0<r<r_0^*$. Since the radius \eqref{r0} is the localization scale of these electron states, they satisfy the long-wavelengths regime if the radius $r_0^*$ much exceeds the lattice constant $a_0$. Thus, the developed theory is applicable to short-range repulsive potentials under the condition $r_0^*\gg a_0$. It should be noted that the discussed bound states are quasi-stationary because they can decay by tunneling through the potential barrier into the continuum of free electron states, as it is pictured in Fig.~2a. As a result, these states acquire energy broadening. If the repulsive potential is strong enough ($\alpha=2\hbar^2/m^\ast_e u_0\ll1$), the previously derived expressions can be applied to describe electron states bound by repulsive potentials in irradiated 2D systems with parabolic electron dispersion (see Ref.~\onlinecite{Kibis_2020}). Namely, the replacements $m_e\rightarrow m^\ast_e$ and $r_0\rightarrow r^\ast_0$ in these expressions yield the energy spectrum of the bound states in irradiated graphene,
\begin{equation}\label{Elm}
\varepsilon_{nm}=\frac{\hbar^2\omega^3\xi^2_{nm}}{4|e|Ev_0}+{\it
O}\left(\alpha\right),
\end{equation}
their energy broadening
\begin{equation}\label{Glm}
{\Gamma}_{nm}=\frac{4\varepsilon_{nm}\alpha^2}{|N^3_m(\xi_{nm})[J_{|m|+1}(\xi_{nm})-J_{|m|-1}(\xi_{nm})]|}+{\it
O}\left(\alpha^3\right),
\end{equation}
and their wave functions
\begin{eqnarray}\label{Flm}
\psi_{nm}&=&\frac{e^{im\varphi}\omega}{\sqrt{\pi}v_0J_{|m|+1}(\xi_{nm})}\left\{
\begin{array}{rl}
J_m\left(\frac{\xi_{nm}\omega r}{v_0}\right), &0<r\le r^\ast_0\\
0, &r\ge r^\ast_0
\end{array}\right.\nonumber\\
&+&{\it O}\left(\alpha\right),
\end{eqnarray}
where $J_m(\xi)$ and $N_m(\xi)$ are the Bessel functions of the
first and second kind, respectively, $\xi_{nm}$ is the $n$th zero
of the Bessel function $J_m(\xi)$ (i.e. $J_m(\xi_{nm})=0$), $n=1,2,3,...$ is the quantum number enumerating zeros of the Bessel function
$J_m(\xi)$, and $m=0,\pm1,\pm2,...$ is the angular momentum. The dependence of the ground state energy $\varepsilon_{10}$ on the field frequency $\omega$ and the field amplitude $E$ is plotted in Fig.~2b. Although the structure of the bound states in electronic systems with the linear and parabolic dispersions is qualitatively similar, the dependence of the states on the field in these systems is substantially different. First, it should be stressed that the bound states exist if the radius of circular trajectory of an electron under the field \eqref{A} much exceeds the crystal lattice spacing, $a_0$. In 2D systems with the parabolic electron dispersion~\cite{Kibis_2020}, this radius is $r_0=|e|E/m_e\omega^2$ and, therefore, the condition $r_0\gg a_0$ can be satisfied in the considered high-frequency limit only for extremely strong field amplitudes $E$. Since such a strong field can destroy a 2D sample, observation of the bound states in 2D systems with the parabolic electron dispersion is experimentally difficult. In graphene, this radius $r^\ast_0$ is defined by Eq.~\eqref{r0} and does not depend on the field amplitude $E$. As a consequence, the condition $r^\ast_0\gg a_0$ can be satisfied for any field amplitude in the broad frequency range due to the giant electron velocity $v_0$. Second, it should be noted that the bound state energy in graphene \eqref{Elm} is $\varepsilon_{nm}\propto 1/E$, whereas this energy in usual 2D systems with the parabolic electron dispersion~\cite{Kibis_2020} is $\varepsilon_{nm}\propto 1/E^2$. Therefore, the bound state energy in graphene under the condition \eqref{c} much exceeds this energy in usual 2D systems. As a consequence, observation of the considered bound states in graphene is most convenient from experimental viewpoint. It follows from Eqs.~\eqref{Elm}--\eqref{Flm}, particularly, that the repulsive potentials confining electron states under a high-frequency field can be considered as quantum dots of new kind if the energy broadening of these states \eqref{Glm} is small enough. In the following, let us discuss possible manifestations of the bound states in electronic transport and optical properties of irradiated graphene.
\begin{figure}[!h]
\includegraphics[width=1.\columnwidth]{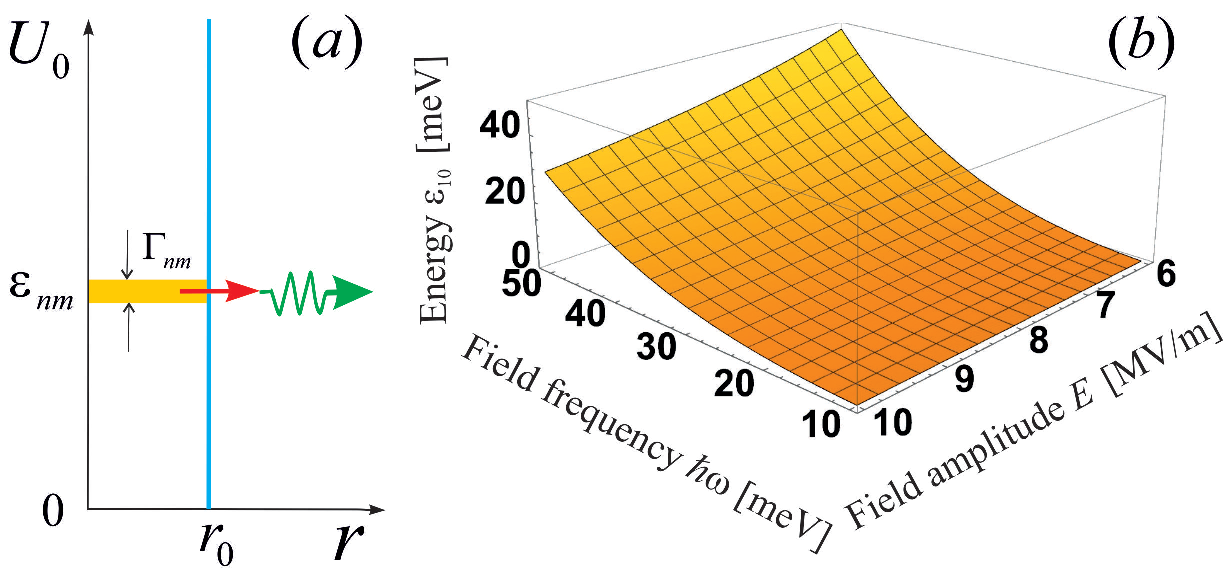}
\caption{The energy spectrum of the electron states localized at a repulsive potential: (a) the tunneling of an electron marked by
the arrow from the bound state with
the energy $\varepsilon_{nm}$ and the broadening $\Gamma_{nm}$ to the state of free conduction electron (the wave  arrow) through
the delta potential barrier
(the vertical line); (b)
the dependence of the ground bound state energy $\varepsilon_{10}$ on the field frequency $\omega$ and the field amplitude $E$. }\label{Fig.2}
\end{figure}

The resistivity of graphene per area, $\rho$, arising from the scattering of conduction electrons trough the quasi-stationary
bound state with the energy $\varepsilon_{nm}$ (the Breit-Wigner resonant scattering) is defined at the zero temperature by the expression~\cite{Kibis_2020}
\begin{equation}\label{tau2}
\rho=\frac{2n_0}{\pi
n_e}\left(\frac{h}{e^2}\right)\frac{(\Gamma_{nm}/2)^2}{(\varepsilon_F-\varepsilon_{nm})^2+(\Gamma_{nm}/2)^2},
\end{equation}
where $h/e^2$
is the resistivity quantum, $n_0$ is the surface density of repulsive scatterers, and $\varepsilon_F$ is the Fermi energy. It follows from Eq.~(\ref{tau2}) that the
resistivity depends on the Fermi
energy $\varepsilon_F$ resonantly. The resonance takes place for
$\varepsilon_F=\varepsilon_{nm}$ and the amplitude of this resonance is
${\rho}_0=(2n_0/\pi n_e)(h/e^2)$. The dependence of the
resistivity (\ref{tau2}) on the field frequency $\omega$ for the ground bound state $\varepsilon_{10}$ is
plotted in Fig.~3 for different electron densities. It
follows from the plot that the resonant amplitude of the resistivity
(\ref{tau2}) is large enough to be observed experimentally,
${\rho}_0\sim\mathrm{\Omega}$, even if the scatterer
density is very small, $n_0\sim10^{-4}n_e$. In the developed theory, the mean free path of conduction electrons, $\ell\sim{n_0}^{-1/2}$, is assumed to be large enough, $\ell\gg r_0^*$. If the mean free path $\ell$  becomes comparable to the radius of the localized electron state $r_0^*$, the electron states confined by the dressed potential become substantially modified. Namely, due to the tunnelling of electrons between the individual confining potentials, the energies of these states become split. Consequently, an electron may hop between the individual confining potentials, which forms the additional channel of electron transport. It should be noted that the localization length of conduction electrons, $\xi$, also depends on the radius $r_0^*$. Indeed, it is described by the known expression $\xi=\sqrt{D\tau_\varphi}$, where $D\sim v\ell$ is the diffusion coefficient of conduction electrons, $v$ is the averaged velocity of electrons, and $\tau_\varphi$ is the coherence time of the electron wave function restricted by inelastic processes. Since $r_0^*$ is the effective radius of an impurity, the increasing of it leads to decreasing the mean free path $\ell$ and, correspondingly, to decreasing the localization length. However, this effect is weak since the localization length $\xi$ normally much exceeds the radius $r_0^*$. From viewpoint of possible superconductivity in irradiated 2D systems, it should be noted that trapping conduction electrons by dressed impurities decreases the density of conduction electrons, which can act on the electron pairing as a destructive factor.
\begin{figure}[!h]
\includegraphics[width=.9\columnwidth]{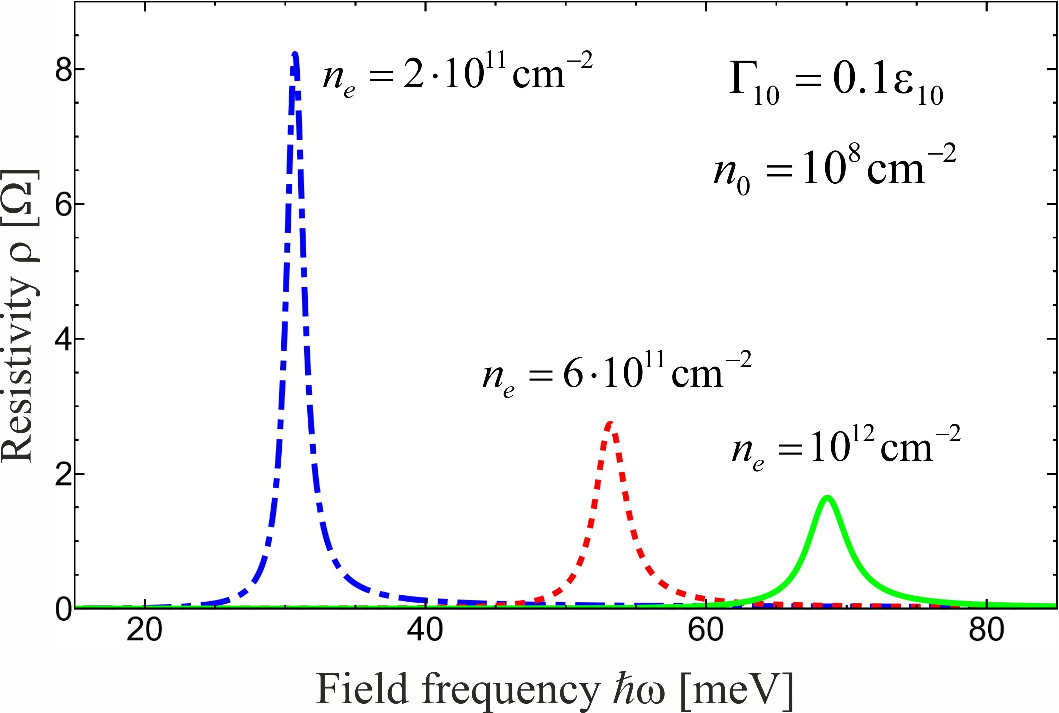}
\caption{Dependence of the resistivity per area, $\rho$, on the
field frequency $\omega$ for the resonant scattering through the ground bound state $\varepsilon_{10}$ with the energy broadening
$\Gamma_{10}=0.1\,\varepsilon_{10}$, scatterer density $n_0=10^{8}$~cm$^{-2}$ and different electron densities
$n_e$.}\label{Fig.3}
\end{figure}

The Kondo effect --- the minimum of electrical resistivity as a function of temperature --- occurs due to the spin interaction between  conduction electrons and electrons bound by magnetic impurities~\cite{kondo1964resistance,de1934electrical,wilson1975renormalization,Wiegmann_1981,Andrei_1983}. However, the spin interaction between confined electron states \eqref{Elm} and free conduction electrons can lead to the Kondo effect as well. When a localized electron stays in its ground state $\varepsilon_{10}$, this interaction can be effectively described by the Hamiltonian~\cite{Iorsh_2021}
\begin{figure}[!h]
\includegraphics[width=.7\columnwidth]{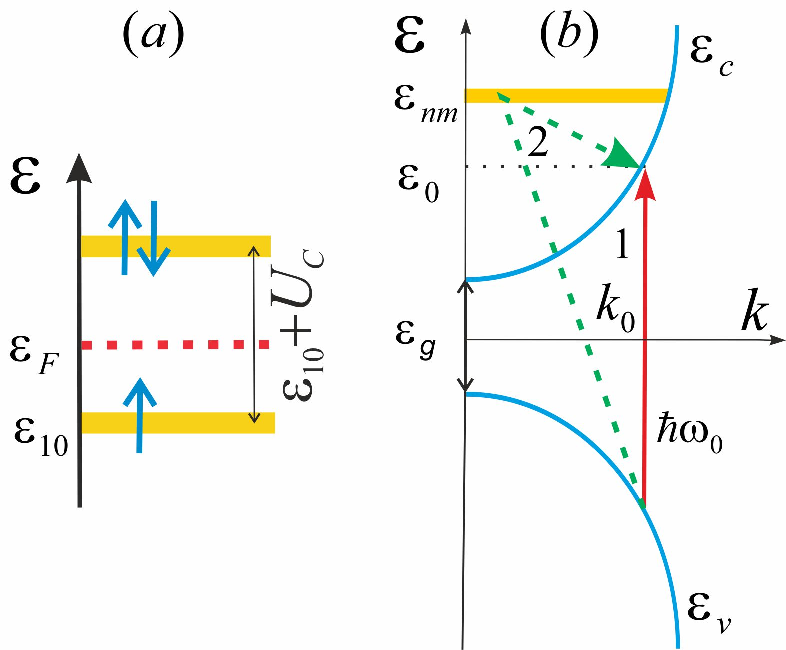}
\caption{(a) The singly-occupied ($\uparrow$) and doubly-occupied ($\uparrow\downarrow$) localized electron state $\varepsilon_{10}$ near the Fermi energy $\varepsilon_F$; (b) Optical transitions from the valence band $\varepsilon_v$ to the
conduction band $\varepsilon_c$ induced by the probing field with
the frequency $\omega_0$: the direct transition (the arrow
$1$) and the transition through the intermediate bound electron state with the energy $\varepsilon_{nm}$ (the arrow
$2$). }\label{Fig.4}
\end{figure}
\begin{eqnarray}
\hat{\cal H}_K&=&\sum_{\mathbf{k},\sigma} (\varepsilon_{\mathbf{k}}-\varepsilon_F)\hat{c}_{\mathbf{k}\sigma}^{\dagger}\hat{c}_{\mathbf{k}\sigma}+\sum_{\sigma}(\varepsilon_{10}-\varepsilon_F)\hat{d}_{\sigma}^{\dagger}\hat{d}_{\sigma}\nonumber\\ &+&U_C\hat{d}_{\uparrow}^{\dagger}\hat{d}_{\uparrow}\hat{d}_{\downarrow}^{\dagger}\hat{d}_{\downarrow}+\sum_{\mathbf{k},\sigma}T_{\mathbf{k}} \left[\hat{c}_{\mathbf{k},\sigma}^{\dagger}\hat{d}_{\sigma}+\mathrm{H.c.}\right], \label{eq:HT}
\end{eqnarray}
where $\varepsilon_{\mathbf{k}}=\hbar^2k^2/2m^\ast_e$ is the energy spectrum of conduction electrons, the symbol $\sigma=\uparrow,\downarrow$ describes the spin orientation,
$\hat{c}_{\mathbf{k},\sigma}^{\dagger}$($\hat{c}_{\mathbf{k},\sigma}$) are the creation (annihilation) operators for delocalized electrons,
$\hat{d}_{\sigma}^{\dagger}$($\hat{d}_{\sigma}$) are the creation (annihilation) operators for localized electrons \eqref{Elm},
\begin{eqnarray}\label{U}
U_C=e^2\int d^2\mathbf{r}\int d^2\mathbf{r}^\prime\frac{|\psi_{10}({r})|^2|\psi_{10}({r^\prime})|^2}{|\mathbf{r}-\mathbf{r}^\prime|} =\frac{\gamma e^2\omega}{\epsilon v_0},
\end{eqnarray}
is the energy of the Coulomb interaction between two electrons with opposite spin directions $\sigma=\uparrow,\downarrow$ occupying the state (\ref{Elm}), $\epsilon$ is the permittivity, $\gamma\approx0.8$ is the numerical constant, and $T_{\mathbf{k}}$ denotes the tunneling matrix element which links the localized electron state (\ref{Elm}) with the delocalized electron state characterized by the
wave vector $\mathbf{k}$. From a physical standpoint, the first component of the Hamiltonian (\ref{eq:HT}) represents the
energy of delocalized electrons, the second component accounts for the energy of electrons  localized in the state (\ref{Elm}), the third component describes the Coulomb energy of the doubly occupied state (\ref{Elm}), and the fourth component is responsible for the tunneling between the localized and delocalized electron states. When tunneling is sufficiently weak, the localized eigenstates of the Hamiltonian (\ref{eq:HT}) correspond to the singly occupied state (\ref{Elm}) with the energy $\varepsilon_{10}-\varepsilon_F$, and the doubly occupied state (\ref{Elm}) with the energy
$2(\varepsilon_{10}-\varepsilon_F)+U_C$. These states are indicated in Fig.~4a by the symbols $\uparrow$ and $\uparrow\downarrow$, respectively. The Kondo effect
emerges due to the spin of a localized electron
and can occur when the singly occupied state is filled,
while the doubly occupied state remains empty. At the zero temperature, this is represented by the conditions $\varepsilon_{10}-\varepsilon_F<0$ and $U_C>\varepsilon_F-\varepsilon_{10}$. Consequently, the irradiation-induced Kondo effect in graphene is feasible within the range of field parameters $E$ and $\omega$  which meet the inequality
\begin{equation}\label{r01}
\sqrt{\frac{\hbar^2v_0\omega \xi_{0}^2}{4|e|E\varepsilon_F}}<\frac{v_0}{\omega}< \frac{\gamma e^2}{2\epsilon\varepsilon_F}+\sqrt{\frac{\hbar^2 v_0\omega\xi_{0}^2}{4|e|E\varepsilon_F}+\left(\frac{\gamma e^2}{2\epsilon\varepsilon_F}\right)^2}.
\end{equation}
Let us stress that the Hamiltonian (\ref{eq:HT}) is equal to the Anderson Hamiltonian which is
fundamental in explaining the microscopic mechanism
of magnetic moment formation in metals~\cite{anderson1961localized}. Therefore, the established Schrieffer-Wolff (SW) unitary transformation~\cite{coleman2015introduction} can be applied to convert the Hamiltonian (\ref{eq:HT}) into the Hamiltonian associated with the Kondo problem~\cite{Andrei_1983}, which defines such a characteristic energy scale of the Kondo effect as the Kondo temperature (see Ref.~\onlinecite{Iorsh_2021} for further details). By substituting $m_e\rightarrow m^\ast_e$ and $r_0\rightarrow r^\ast_0$ into the previously developed Kondo problem theory
for an irradiated 2D system with the parabolic electron dispersion~\cite{Iorsh_2021}, one can find that the Kondo temperature in graphene is of several Kelvin for the field \eqref{A} with the reasonable parameters $\hbar\omega\sim\mathrm{meV}$ and $E\sim\mathrm{MV/m}$. Thus, the Kondo effect in irradiated graphene can be observable using state-of-the-at experimental techniques.

To detect the bound electron states \eqref{Elm} by optical methods, graphene should be irradiated by an electromagnetic field consisting of the two modes: The first mode is the strong off-resonant field \eqref{A} which yields the bound states, whereas the second one is the weak probing resonant field with the amplitude $E_0$ and the frequency $\omega_0$, which generates the electron transitions between the valence and conduction bands of graphene. Therefore, both the direct interband electron transitions (marked by the solid arrow $1$ in Fig.~4b) and the interband electron transitions through intermediate bound states (marked by the dashed arrow $2$ in Fig.~4b) take place. To describe the low-energy optical properties, it is enough to take into account only the ground
bound state $\varepsilon_{10}$. Assuming the scatterer density to be small enough, $n_0{r_0^*}^2\ll 1$, the intensity of the probing field absorption, which arises from these two types of optical transitions, is described by the expression~\cite{Kibis_2021_1}
\begin{align}\label{IF}
&I=\left(\frac{\omega_0 m^\ast_e|D_{cv}|^2E_0^2}{2\hbar^2}\right)
\Bigg[1+\frac{\hbar^2n^2_0{\Gamma_{10}}\Phi^2_{10}(k_0)}
{m^\ast_e[(\varepsilon_{0}-\varepsilon_{10})^2+(\Gamma_{10}/2)^2]}\nonumber\\
&-\left.\frac{{2}n_0\hbar\Gamma^{1/2}_{10}\Phi_{10}(k_0)(\varepsilon_{0}-\varepsilon_{10})}
{\sqrt{m^\ast_e}\,[(\varepsilon_{0}-\varepsilon_{10})^2+(\Gamma_{10}/2)^2]}\right],
\end{align}
where $D_{cv}$ is the interband matrix element of dipole moment,
$k_0=\sqrt{(\hbar\omega_0-\varepsilon_g)m^\ast_e}/\hbar$
is the electron wave vector corresponding to the direct
interband electron transition,
$\varepsilon_0=\hbar^2k^2_0/2m^\ast_e$ is
the electron energy corresponding to this wave vector, and $\Phi_{10}({k})=\int d^2\mathbf{r}\,\psi_{10}(r)e^{i\mathbf{kr}}$ is the Fourier transform of the
ground bound state wave function. The first term in the square brackets of Eq.~(\ref{IF}) arises
from the direct optical transition. Correspondingly, the intensity of direct absorption of
the probing field is ${I}_0={\omega_0
m^\ast_e|D_{cv}|^2E_0^2}/{2\hbar^2}$. The second term there arises from the
optical absorption through the bound state with the energy $\varepsilon_{10}$, whereas the third term
originates from the quantum interference of these two absorption ways. Since the interference term
changes its sign at the energy
$\varepsilon_0=\varepsilon_{10}$, we arrive at the
asymmetrical structure of the absorption spectrum plotted in
Fig.~5, which is typical for the Fano resonances~\cite{Fano_1961} and can be easily detected experimentally.
\begin{figure}[!h]
\includegraphics[width=1.\columnwidth]{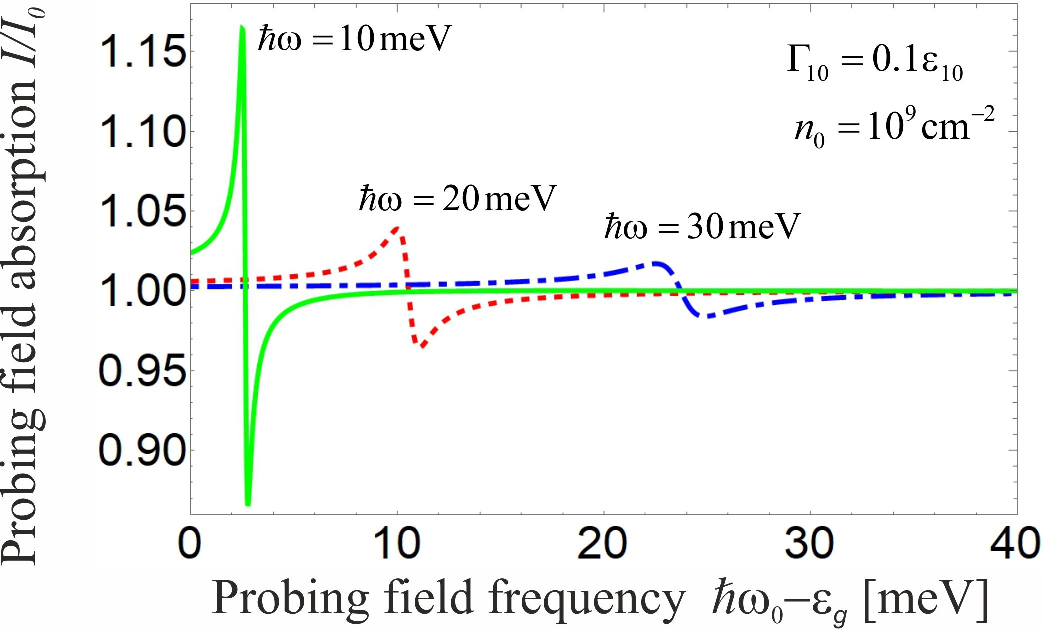}
\caption{The intensity of the probing field absorption, $I$, as a function of the probing field frequency, $\omega_0$, for the density of scatterers $n_0=10^9$~cm$^{-2}$, the energy broadening $\Gamma_{10}=0.1\varepsilon_{10}$, and different frequencies of the dressing field, $\omega$.}\label{Fig.5}
\end{figure}

\section{Conclusion} We demonstrated theoretically that a circularly polarized irradiation of graphene induces a local minimum of a repulsive potential, which results in bound electron states localized near the minimum. It is shown that the observability conditions for these states in graphene can be realized substantially easier than in usual 2D systems with the parabolic electron dispersion. The states can manifest themselves, particularly, in the Breit-Wigner resonant scattering of conduction electrons through them, the Kondo effect arisen from electrons localized at the states, and the asymmetrical structure of optical spectra originated from the Fano resonance of light absorption. Since these phenomena can be observed in graphene for experimentally achievable field parameters, the field-induced electron states bound by repulsive potentials can be detected in state-of-the-art measurements.

\begin{acknowledgments}
The reported study was funded by the Russian Science Foundation (project 20-12-00001).
\end{acknowledgments}

\section*{Data availability statement}
All data that support the findings of this study are included
within the article (and any supplementary files).

\end{document}